\documentclass[aps,prb,preprint,showpacs]{revtex4}

\usepackage{epsf}

\usepackage{amssymb}

\begin{document}


\title{The exchange coupling between the valence electrons of the fullerene cage and the electrons of the N atoms in N@C$_{60}^{-1,3}$}
\author{L. Udvardi }
\affiliation{Budapest University of Technology and Economics, Department of Theoretical Physics,
 Budafoki \'ut 8, H-1111 Budapest, Hungary}

\begin{abstract}
MCSCF calculations are performed in order to determine the exchange coupling  
between the 2p electrons of the N atom and the LUMOs of the fullerene cage in the
case of mono- and tri-anions of N@C$_{60}$. The exchange coupling resulted by our 
calculations is large compared to the hyperfine interaction. The strong coupling can 
explain the missing EPR signal of the nitrogen in paramagnetic anions.
\end{abstract}

\pacs{31.15.Ar}
\maketitle

\section{Introduction}
Since the discovery of the first endohedral fullerene a lot of interests have been
attracted by this area of the nanotechnology. 
Many metal atom can be encapsulated by using 
discharge techniques or ion implantation.  In all the cases the metal atom interacts
strongly with the fullerene and acts as an electron donor occupying an 'off-centered'
position inside the cage. In contrast the nitrogen in N@C$_{60}$  is situated at the center 
of the molecule and retains its S=3/2 spin quartet atomic state \cite{waib}.  
This amazing property of the encapsulated N atom  triggered several research on its
 possible application in  quantum computing and spin labeling.
Several publications \cite{qc1,qc2,qc3,qc4} studied the promise and limitations of using 
endohedral fullerenes as quantum information carriers. Mehring {\it et al.} \cite{qc5} 
recently pointed out experimentally the entanglement of the nuclear spin and the electronic 
spin of the encaged N atom. 

The changes of the characteristic EPR signal of the quartet electronic spin of the N atom
makes it an ideal probe for monitoring chemical reactions of C60 \cite{spinl}. 
 During the last decade a great deal of excitement has been brought by the discovery of the
superconductivity of the alkali-doped fullerenes. In this type of fullerene compounds
the valence electrons of the ionized alkali atoms partially occupy the bands formed by 
the LUMOs of the C$_{60}$ molecules. The applicability of the quartet atomic state of the N 
atom as a spin label  depends on the strength of the interaction between the 2p electrons
of the N atom and the valence electron of the fullerene cage. An interaction which is small
compared to the hyper-fine interaction, results in a line width effect of the EPR 
signal and the N@C$_{60}$ is a good candidate for a spin labeling agent. 
In the case of strong coupling
the EPR signal of the system is completely changed and the lines corresponding to the
valence electrons of the N atom  are hard to
identify in the signal of the paramagnetic system.  

 The interaction between the 2p electrons of the N atom and the valence electrons of the
C$_{60}$ can be described by a Heisenberg like effective Hamiltonian 
$H_{int} = J\mathbf{S}_{N}\mathbf{S}_{C_{60}}$ where $\mathbf{S}_{N}$ and 
 $\mathbf{S}_{C_{60}}$ denote the spin operator for the valence electrons of
the N atom and the the C$_{60}$, respectively, and J is the exchange coupling 
characterizing the strength of the interaction. 
The aim of the present paper is to determine theoretically the exchange
coupling of the effective Hamiltonian. 
The exchange coupling has importance not only for EPR measurements but it plays essential
 role in the description of the transport through  magnetic molecules \cite{transport}
 which is particularly interesting from the point of view of spintronics.

\section{Computational details}
 The calculations have been performed using the Gamess quantum chemical program package  \cite{gamess}. The proper description of the open-shell N@C$_{60}^{-1}$ and N@C$_{60}^{-3}$  anions requires
multi-determinant wave functions.  The restricted open-shell (ROHF) calculations  in the
Gamess package are accessible via the generalized valence bond (GVB) or  
the multi-configurational self-consistent field (MCSCF) methods using an appropriate  active space.
The energy of the anions of N@C$_{60}$ with different multiplicity has been determined by 
means of CAS SCF calculations where the active space is confined to the {\it 2p} orbitals of the nitrogen
atom and the three fold degenerate LUMOs of the fullerene molecule.  For the clear interpretation of the 
results the excitations from the orbitals of the nitrogen to the LUMOs of the cage, and vice versa,
were excluded from the active space applying the occupation restricted multiple active space 
\cite{ormas} technique. The MCSCF treatment of the open-shell systems using such a small active space
is practically equivalent to the ROHF level of calculations. The  calculation has been performed using split valence 631g basis on the carbon atoms.
For the better description of the week interaction between the encapsulated atom 
and the fullerene molecule \cite{jcp} the basis on the N atom is extended by 
additional diffuse {\it p} orbitals and two d polarization functions (631+g(dd)).     
It is well known that in order to describe the electronic structure of 
negatively charged species application of diffuse basis functions is necessary. 
In our case the excess charge is distributed uniformly among the 60 carbon atoms 
and the lack of the diffuse basis on the carbon atoms  does not affect dramatically
our results. However, in order to check the sensitivity of the exchange coupling to
the applied basis the calculations have also been performed with the Dunning's 
double zeta \cite{dh} and the split valence 631+g basis sets on the carbon atoms.
The geometry of the N@C$_{60}^{-1}$ molecule  in the S=1 state  and the 
N@C$_{60}^{-3}$  in the high spin S=3 state have been optimized at ROHF level and
it is retained  during the calculations of the energy of the systems with 
different multiplicity. 

\section{Results and discussions}
It has been shown experimentally that in the highly reduced  states of the 
N@C$_{60}$  the excess electrons occupy the LUMOs of the fullerene and the N atom 
inside the cage remains in spin quartet state \cite{anion}. In order to check the 
consistency of  our calculations to the experimental findings  we performed a set 
of ROHF calculations on the mono- and tri-anions populating at first the {\it 2p}
orbitals of the nitrogen and then populating the LUMOs of the C$_{60}$. The results
are summarized in Table \ref{tbl:eanion}. The valence electrons of the nitrogen 
referred as $N2p$ in Table \ref{tbl:eanion}  occupy the $7t_{1u}$ orbitals of the 
endohedral complex between the $6h_u$ HOMO and $8t_{1u}$ LUMO of the C$_{60}$ in 
agreement with the result of ref \cite{Lu}. Rather different value for the 
one-electron energy of the $N2p$ orbitals is reported by Greer \cite{Greer}. 
This discrepancy is originated from the different treatment of the open-shell 
problem as it is discussed in ref. \cite{Plakhutin}.
In the case of the mono-anion  the energy of the two triplet states were compared
while in the case of the triply ionized molecule the energy of the singlet state 
with fully occupied valence orbitals of N was compared  to the high spin state of 
the N@C$_{60}^{-3}$.
For both ions the  system with intact N atom were energetically more favorable in 
agreement with the EPR measurements \cite{anion}.

The interaction between the electrons of the nitrogen atom and the valence electrons
on the C$_{60}$ anion is described by a Heisenberg-like effective Hamiltonian:
\begin{equation}
 H_{int} = J\mathbf{S}_N\mathbf{S}_{c_{60}} 
\end{equation}
where $J$ is the coupling constant, $\mathbf{S}_N$ and $\mathbf{S}_{c_{60}}$ 
are the spin of the nitrogen atom and the C$_{60}$ anion, respectively.   
The square of the total spin operator 
$\mathbf{S}^2 = (\mathbf{S}_N+\mathbf{S}_{c_{60}})^2$ 
commutes with the Hamiltonian of the full system 
$H = H_{N} + H_{C_{60}} +  H_{int}$ and ,consequently, its eigenvalue is a good 
quantum number.  Expressing the interaction  in terms of the spin of the
subsystems and the spin of the whole molecule:
\begin{equation}
 H_{int} = \frac{1}{2}J\left (\mathbf{S}^2 - \mathbf{S}_N^2 - \mathbf{S}_{c_{60}}^2 \right )
\end{equation}
the energy can be simply given as:
\begin{equation} \label{eq:ES}
 E_S = E_0 + \frac{1}{2}JS(S+1) \;\;,
\end{equation}
where $E_0$ denotes the energy of the separated systems and the subscript $S$ 
indicates the explicit dependence of the energy on the multiplicity.
Since our interaction Hamiltonian can describe only such processes in which 
$\mathbf{S}_N$ and $\mathbf{S}_{c_{60}}$ are unchanged the excitations altering 
the spin of the subsystems, namely the hole and the particle are on different 
species, has to be excluded from the configuration space.
In the case of N@C$_{60}^{-1}$ $S_N = 3/2$ and $S_{C_{60}} = 1/2$ spanning an 
8 dimensional direct product space. The  total spin can have the values of $S=1$ 
or $S=2$ with the corresponding energies 
\begin{equation}
 E_{S=1} = E_0 +  J  \;, \;\;\;\;\;
 E_{S=2} = E_0 + 3J \;.
\end{equation} 
Comparing the energy of the triplet and quintet state one can easily extract the 
exchange coupling as:
\begin{equation}
 J = \frac{1}{2}\left (  E_{S=2} - E_{S=1}  \right )
\end{equation} 

The results of the MCSCF calculations using 631g and DH basis are summarized in 
Table \ref{tbl:c60n-1}.  Although the application of the double zeta basis resulted 
in considerably deeper total energy the deviation of the exchange couplings is small.

\begin{figure}
 \begin{center}
 \epsfxsize=10cm
 \epsfbox{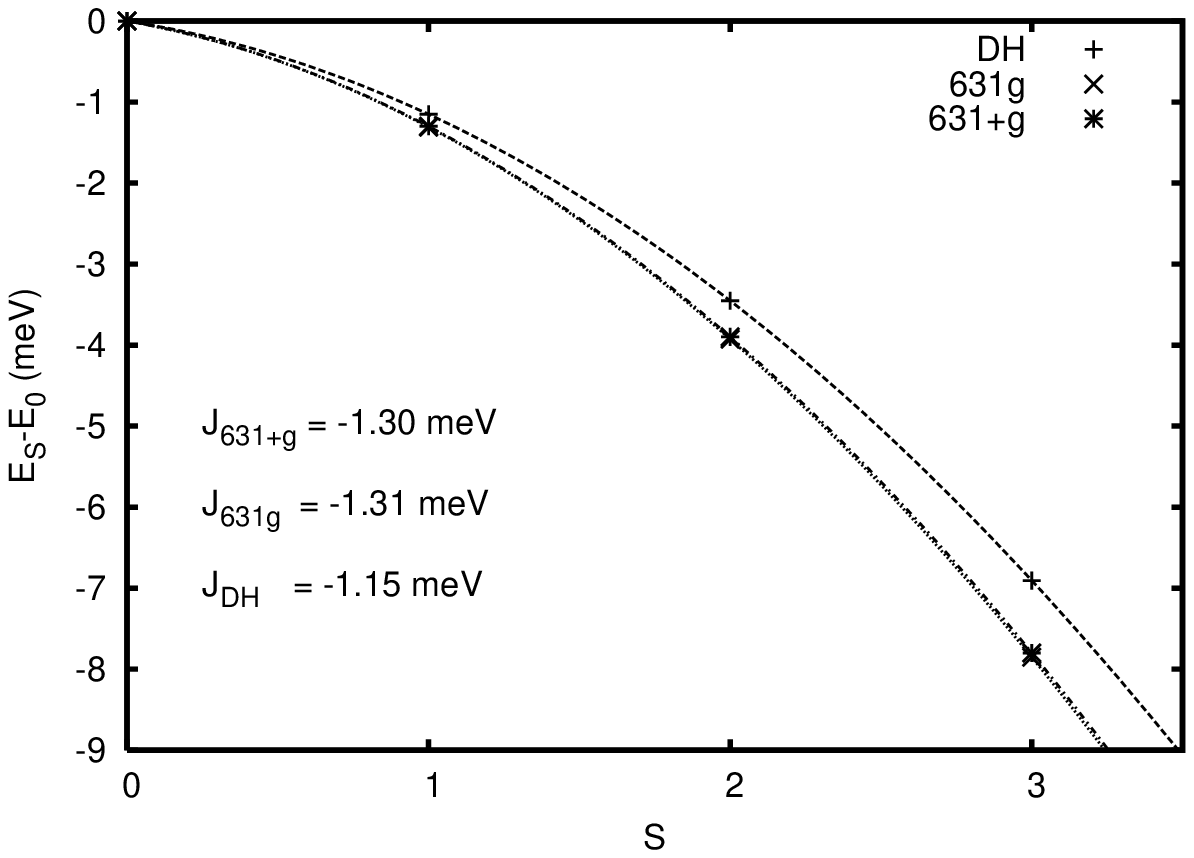}
 \caption{\label{fig:ec60n-3} Energies of N@C$_{60}^{-3}$ corresponding to different multiplicity and the parabola fitted to the data points. The energy $E_0$ independent of spin is subtracted. }
 \end{center}
\end{figure}

In the case of the triply ionized N@C$_{60}$ the valence electrons form a 
$S_{C_{60}} = 3/2$ state on the LUMOs of the fullerene molecule according to the 
Hund's rule. From the two quartet states, $\mathbf{S}_N$, $\mathbf{S}_{C_{60}}$, 
four eigenstate of the $\mathbf{S}^2$ operator can be constructed with the spin of 
$S=0,1,2,3$, respectively. The corresponding energies as a function of 
$S$ must be on a parabola according to  Eq. \ref{eq:ES}.  The results provided by
the MCSCF calculations using three different basis sets are shown
by Fig. \ref{fig:ec60n-3}. The energies can nicely be fitted by the parabola given 
by Eq. \ref{eq:ES}.  The exchange couplings obtained by using the split valence 
basis with and without diffuse {\it p} orbitals are practically the same. 
The inclusion of the diffuse basis functions on the carbon atoms resulted in 
negligible change.  Although the magnitude of the exchange coupling corresponding 
to the double zeta basis is somewhat smaller than those provided by the split 
valence basis the agreement between them is satisfactory.  

Ferromagnetic exchange couplings between the $2p$ orbitals of the N atom and the 
valence electrons of the fullerene molecule have been found in both anions. 
The exchange coupling of approximately 1~meV provided by our calculations for both
systems is within the range of those found in organic ferromagnets \cite{metal}. 
This relatively strong coupling between the valence electrons of the nitrogen and 
the valence electrons of the fullerene cage could be responsible for the 
disappearance of the nitrogen lines in the EPR spectrum of N@C$_{60}$ anions with
partially filled LUMOs \cite{anion}.

\section{Conclusions}
 In conclusion, ROHF and MCSF calculations have been performed on singly and triply 
ionized anions of N@C$_{60}$ in order to determine the effective exchange coupling 
between the valence electrons of the encapsulated N atom and the fullerene cage. 
In agreement with experiments we found that the excess electrons occupy the LUMOs 
of the fullerene molecule and the entrapped atom keeps its atomic character. 
The interaction between the valence electrons of the N atom and the LUMOs of the 
C$_{60}$ can be well described by a Heisenberg like Hamiltonian. The size of the 
exchange couplings obtained by our calculations are much larger then the hyperfine 
interaction and can explain the results of EPR measurements on radical anions of 
N@C$_{60}$.

\section{Acknowlegments}
This work is supported by the Hungarian National Science Foundation 
(contracts OTKA T038191 and T037856).

\newpage
\begin{table}
  \begin{center}
  \begin{tabular}{clllc}\hline\hline
    \            & Configuration                        & E$_{total}$(Hartree) & $\Delta$ E (eV)  \\ \hline
 N@C$_{60}^{-1}$ & N$2p^4$C$_{60}8t_{1u}^0$ S = 1 & -2325.30365 &   \   & {\it (a)}  \\ 
    \            & N$2p^3$C$_{60}8t_{1u}^1$ S = 1 & -2325.37155 & -1.84 & {\it (b)}  \\ \hline
 N@C$_{60}^{-3}$ & N$2p^6$C$_{60}8t_{1u}^0$ S = 0 & -2324.74514 &   \   & {\it (a)}  \\ 
    \            & N$2p^3$C$_{60}8t_{1u}^3$ S = 3 & -2325.10707 & -9.84 & {\it (b)}  \\ \hline \hline
  \end{tabular} 
 \caption{ \label{tbl:eanion} Energies of N@C$_{60}^{-1}$ and  N@C$_{60}^{-3}$ with
          excess electron(s) occupying the $2p$ orbitals of the N atom ({\it a}) and 
          the LUMOs of the C$_{60}$ molecule ({\it b}).} \end{center}
\end{table}   

\begin{table}
 \begin{center}
 \begin{tabular}{cccc} \hline\hline
  basis & E$_{S=1}$ (Hartree) & E$_{S=2}$ (Hartree)  &   J (meV) \\ \hline
  631g  &      -2325.371558       &  -2325.371673            &  -1.56 \\ 
  DH     &      -2325.515025       &  -2325.515134            &  -1.49 \\ \hline\hline
\end{tabular} 
 \end{center}
\caption{\label{tbl:c60n-1} Energy of the N@C$_{60}^{-1}$ resulted by MCSCF calculations using
 split valence (631g) and double zeta (DH) basis on the carbon atoms and the exchange coupling 
 extracted from the energies. }
\end{table}

\newpage

\begin{thebibliography}{00}

\bibitem{waib}
T.Almeida Murphy, T. Pawlik, A. Weidinger, M. H\"ohne, R. Alcala, J.-M. Spaeth,
Phys.Rev. Letter. {\bf 77} (1996) 1075
B. Pietzak, M. Waiblinger, T.Almeida Murphy, A. Weidinger, M. H\"ohne, E. Dietel,
A. Hirsch, Chem. Phys. Letters {\bf 279} (1997) 259
\bibitem{qc1}
W. Harneit, Phys. Rev. {\bf A 65} (2002) 032322 
\bibitem{qc2}
D. Suter and K. Lim, Phys. Rev. {\bf A 65} (2002) 052309
\bibitem{qc3}
J. Twamley, Phys. Rev. {\bf A 67} (2003) 052318 
\bibitem{qc4}
M. Feng and J. Twamley, Phys. Rev. {\bf A 70} (2004) 032318 
\bibitem{qc5}
 M. Mehring, W. Scherer, and A. Weidinger, Phys. Rev. Lett. {\bf 93} (2004) 206603 
\bibitem{spinl}
E. Dietel, A. Hirsch, B. Pietzak, M. Waiblinger, K. Lips, A. Weidinger, A. Gruss, 
K.P. Dinse, J. Am. Chem. Soc.  {\bf 121} (1999) 2432
\bibitem{transport}
F. Elste and C. Timm, Phys. Rev. {\bf B 71} (2005) 155403 

\bibitem{gamess}
 M.W.Schmidt, K.K.Baldridge, J.A.Boatz, S.T.Elbert,
M.S.Gordon,  J.H.Jensen, S.Koseki, N.Matsunaga,
K.A.Nguyen, S.J.Su, T.L.Windus, M.Dupuis, \\  J.A.Montgomery
J.Comput.Chem.  {\bf 14}, 1347-1363(1993)
\bibitem{ormas}
J.Ivanic  J.Chem.Phys.  {\bf 119} ((2003) 9364, 9377
\bibitem{jcp}
J.M. Park, P. Tarakeshwar, and K.S. Kim J. Chem. Phys.{\bf 116} (2002) 10684
\bibitem{dh}
T.H.Dunning, Jr., P.J.Hay Chapter 1 in "Methods of Electronic Structure Theory", H.F.Shaefer III, Ed. Plenum Press, N.Y. 1977, pp 1-27.
\bibitem{anion}
KP. Dinse,  B. Godde,  P. Jakes, M. Waiblinger, A. Weidinger, A. Hirsch,   
Abstr. Pap. - Am. Chem. Soc.  (2001)  221st  IEC-199.  CODEN: ACSRAL  ISSN: 0065-7727.,
P. Jakes, B. Godde, M. Waiblinger, N. Weiden, K.P. Dinse, A. Weidinger,  
AIP Conference Proceedings  {\bf 544} (2000) 174 

\bibitem{metal}
D.A. Shultz,K.E. Vostrikova, S.H. Bodnar, Hyun-Joo Koo,
Myung-Hwan Whangbo, M.L. Kirk, E.C. Depperman, and J.W. Kampf
J. Am. Chem. Soc.  {\bf 125} (2003) 1607 

\bibitem{Lu}
J. Lu, X. Zhang, X. Zhao, Chem. Phys. Lett. {\bf 312} (1999) 85

\bibitem{Greer}
J.C. Greer,  Chem. Phys. Lett. {\bf 326} (2000) 567 

\bibitem{Plakhutin}
B.N. Plakhutin, N.N. Breslavskaya, E.V. Gorelik, A.V. Arbuznikov, 
Journal of Molecular Structure: THEOCHEM {\bf 727} (2005) 149 

\end{thebibliography}
\end{document}